\documentclass[aps,prl,showpacs,twocolumn,groupedaddress]{revtex4-1}
\usepackage{graphicx,graphics,color,epsfig}
\usepackage{amsmath}
\usepackage{amssymb}
\usepackage{amsfonts}
\usepackage{amsfonts}
\usepackage{epstopdf}
\usepackage{bm}
\usepackage{times,xspace}
\usepackage{color}

\begin{document}

\title{Interaction induced staggered spin-orbit order in two-dimensional electron gas}

\author{Tanmoy Das\\
Theoretical Division, Los Alamos National Laboratory, Los Alamos, NM 87545, USA.}

\begin{abstract}
We propose and formulate an interaction induced staggered spin-orbit order as a new emergent phase of two-dimensional Fermi gases. We show that when some form of inherent spin-splitting via Rashba-type spin-orbit coupling renders two helical Fermi surfaces to become significantly `nested', a Fermi surface instability arises. To lift this degeneracy, a spontaneous symmetry breaking spin-orbit density wave develops, causing a surprisingly large quasiparticle gapping with chiral electronic states. Since the staggered spin-orbit order is associated with a condensation energy, quantified by the gap value, destroying such spin-orbit interaction costs sufficiently large perturbation field or temperature or de-phasing time. BiAg$_2$ surface state is shown to be a representative system for realizing such novel spin-orbit interaction with tunable and large strength, and the spin-splitting is decoupled from charge excitations. These functional properties are relevant for spin-electronics, spin-caloritronics, and spin-Hall effect applications.
\end{abstract}


\maketitle \narrowtext

In semiconductor heterostructures, a charge particle moving on a symmetry-breaking electric field experiences an effective `anisotropic' magnetic field due to relativistic effect, which couples to its spin. Such spin-orbit coupling (SOC), known as Rashba-\cite{Rashba,Bychkov} or Dresselhaus-type\cite{Dresselhaus} SOC, has proven to be a useful ingredient for realizing many physical concepts such as spin Hall effect,\cite{SHall} spin torque current\cite{STorque08,STorque10}, spin domain reversal,\cite{SReversal} and exotic emergent superconducting,\cite{SSC} and magnetic phases,\cite{SMag} which are relevant for the applications of spin-electronics,\cite{spintronics} spin-caloritronics\cite{caloritronics} and quantum information processing. Along with long spin lifetime, spin relaxation, and immunity to perturbations or disorders, a highly desired recipe for these applications is decoupling the spin current from associated charge flow. This is a difficult task as the electric field generated either by the broken inversion symmetry and/ or by the magnetic field itself naturally generates charge current perpendicular to the spin current, thus greatly hinders the spin transport.\cite{SSeparation,SPure} Despite the discovery of giant SOC in a large class of condensed matter systems,\cite{Au111,BiAg,BiTeI,coldatom,TI,BiAg2PRB,BiAg2PRL} practical realization and device implementation of these concepts have thus posed challenging.\cite{spintronics,caloritronics}

In this work, we introduce a theoretical proposal for generating and manipulating staggered spin-orbit entangled helical state via electronic interaction. A key ingredient for realizing such state is to have some form of intrinsic SOC, whose strength is not important, prior to the inclusion of interaction. For such systems, the ground state is defined by more exotic quantum numbers such as total angular momentum ($J$ for $j-j$-type SOC) or helical quantum number ($\nu=\pm$ for Rashba- or Dresselhaus-type SOC), rather than typical spin-, orbital-, or momentum alone. Therefore, if the interaction strength is less than the SOC strength, a typical spin- or orbital-density wave alone is prohibited to form. On the other hand, we demonstrate here that novel emergent phases of matter can arise in the spin-orbit channel even without necessarily breaking the time-reversal symmetry.

We formulate the aforementioned postulates on the basis of a Rashba-type SOC ground state; however the idea is general and can be extended to other systems. Rashba SOC splits the non-interacting FS into two concentric helical Fermi pockets, and thereby a FS `hot-spot' ${\bm Q}$ develops where degeneracy induces FS nesting between the opposite helical states. As a result of such instability, a translational symmetry breaking spontaneous collective ordering of helical degree of freedom develops, causing a giant SODW. The resulting SODW spatially modulates with a periodicity determined by the `hot-spot' wavevector ${\bm Q}$. The SODW renders gapping in the quasiparticle states, and the corresponding gap energy $\Delta$ thermodynamically represents its strength. Unlike in topological insulators or in other SOC systems where a spin-degenerate point exists with {\it zero} gapping, here the finite gap $\Delta$ protects the SODW phase from external perturbations and spin de-phasing. This means the spin-splitting survives up to a finite strength of Zeemen energy $E_Z$ (comparable to $\Delta$), determining the critical value of a time-reversal breaking perturbation such as magnetic field, $B_c$, and spin de-phasing time $\tau_s$. As a proposal, we show that a feedback effect of this order is the presence of a spin-orbit collective mode, which physically represents the interaction between two electrons with opposite spins and orbitals, or spin-orbit entangled quantum numbers, and that it can be detected via two-particle probes such as polarized inelastic neutron scattering.

\begin{figure}[t]
%\hspace{0.cm}
\rotatebox[origin=c]{0}{\includegraphics[width=1.0\columnwidth]{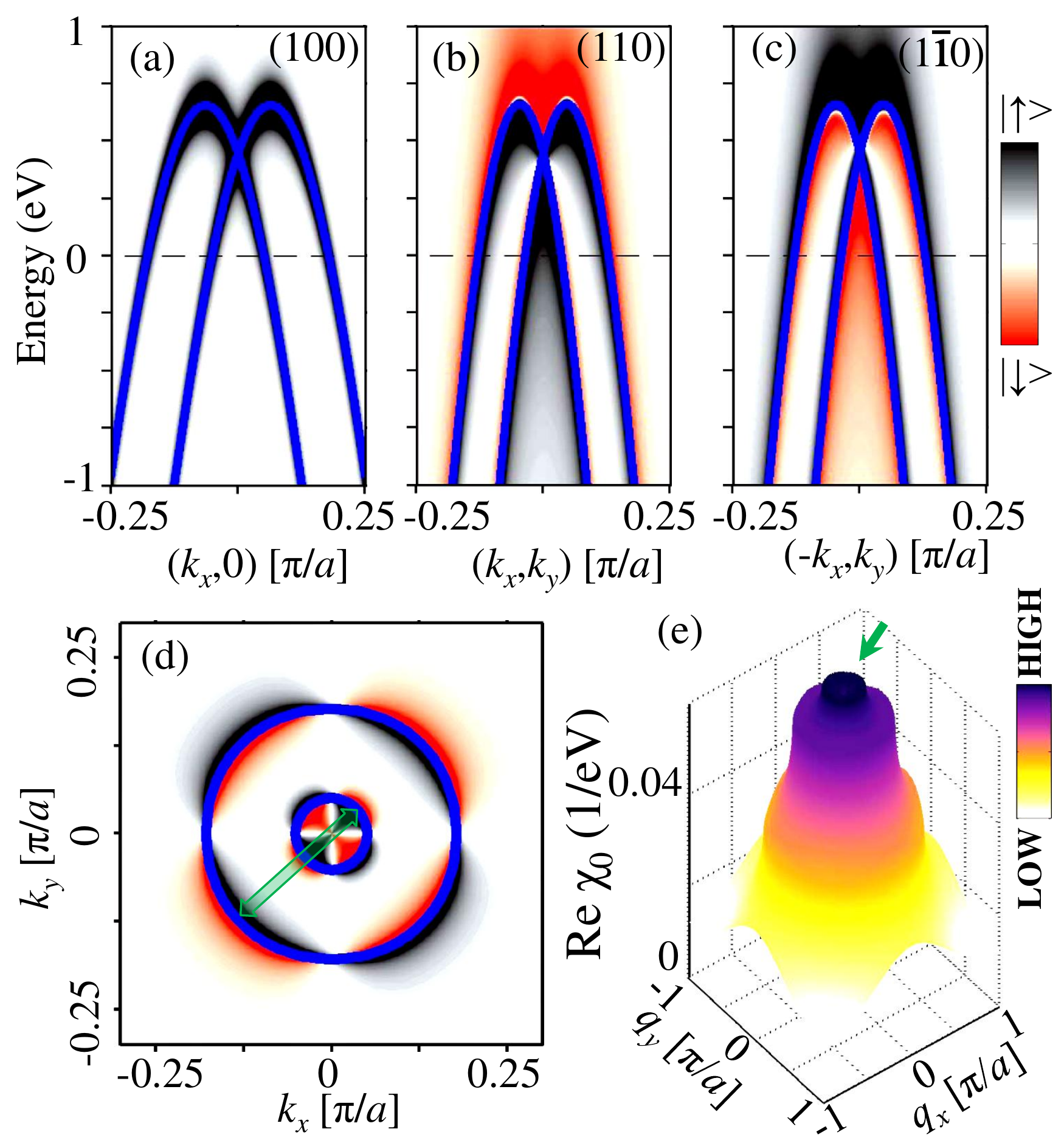}}
\caption{(Color online) ({a}) Non-interacting bands split by single-fermion Rashba-type SOC plotted along (100)-direction. The blue solid line is the coherent eigenstate, while the background is the associated single-particle spin-resolved spectral weight. The spin-polarization is depicted by a red (down-spin) to black (up-spin) gradient colormap. Along (100), the Rashba term vanishes, and thus the spin-resolved spectral weight gradient is absent here. The counter polarization of the spin texture along (110)- and (1$\bar{1}$0)-directions is demonstrated in ({b}) and ({c}), respectively. ({d}) Corresponding FS map is plotted in the same color scale. ({e}) Real part of non-interacting susceptibility at zero excitation energy reveals the development of paramount inter-helical state nesting at the `hot-spot' vector ${\bm Q}\sim0.115(\pi,\pi)$.}
\label{fig1}
\end{figure}

{\it Lattice model for Rashba coupling:-}We start with a system of two-component Fermi gas in the presence of Rashba-type SOC. In the single-particle description of the system, the non-interacting FS is spin-polarized on the two-dimensional (2D) momentum space. For some systems, as in the surface state of BiAg$_2$ deposited on various substrates,\cite{BiAg2PRB,BiAg2PRL} the FSs can yield a shape to generate dominant FS nesting, and hence an unstable one-body ground state. If two momenta across the nesting `hot-spot' is connected by spin flips, it cannot support a charge density wave scenario. Furthermore, if the nesting commences between different segments of the same band, a spin-density wave may arise if the interaction strength can overcome the SOC strength. On the other hand, in such case a SODW can arise via inter-helical FS nesting instability even if the interaction strength is lower than the SOC strength.

The non-interacting Hamiltonian in the two component fermion fields $\psi_{\bm k}=[\psi_{{\bm k},\uparrow}, \psi_{{\bm k},\downarrow}]^{T}$ as $H_0({\bm k}) = \psi_{\bm k}^{\dag}[\xi_{\bm k}{\bf 1} -i\alpha_R({\sigma}_x\sin{k_y}-{\sigma}_y\sin{k_x})]\psi_{\bm k}$. Here  $\xi_{\bm k}$ is the free-fermion dispersion term, modeled by nearest-neighbor electronic hopping $t$ as  $\xi_{\bm k}=-2t[\cos{k_x}+\cos{k_y}]-E_F$, where $E_F$ is the chemical potential. The second term is the 2D lattice generalization of the standard Rashba SOC term $H_{so}=-i\alpha_R(\hat{\bm \sigma}\times\hat{\bm k})_z$, with $\hat{\bm \sigma}$ being the Pauli matrices and $\alpha_R$ being the Rashba SOC strength. The helical dispersion spectrum of $H_0$ is $E_{{\bm k},0}^{\pm}=\xi_{\bm k}\pm\alpha_R[\sin^2{k_x}+\sin^2{k_y}]^{1/2}$. The parameters are obtained by fitting to the experimental dispersion from Ref.~\onlinecite{BiAg2PRL}, as listed in Ref.~\onlinecite{parameters} and the computed dispersion is plotted in Figs.~\ref{fig1}({a}), \ref{fig1}({b}), \ref{fig1}({c}) along different high-symmetry lines, and the corresponding FS is given in Fig.~\ref{fig1}(d).

{\it FS instability:-}We now investigate the FS instabilities of the system by evaluating the bare susceptibility $\chi$ in the particle-hole channel. $\chi$ is computed by convolving the single-particle Green's function $G({\bm k},i\omega_n)=(i\omega_n-H_0({\bm k}))^{-1}$, yielding $\chi({\bm q},ip_m)=\sum_{\bm k,n}G({\bm k},i\omega_n)G({\bm k}+{\bm q},i\omega_n+ip_m)$, where $i\omega_n$ and $ip_m$ are the fermionic and bosonic Matsubara frequencies, respectively. The result plotted in the 2D ${\bm q}$ space at zero energy (the real frequency is obtained by taking analytical continuation from the Matsubara frequency) in Fig.~\ref{fig1}(e) exposes that the nesting at ${\bm Q}=0.115(\pi,\pi)$ is dominant. ${\bm Q}$ connects two momenta lying on different helical bands as shown in Fig.~\ref{fig1}({d}). Due to the definite chirality of the FS, different orientations of ${\bm Q}=0.115(\pm\pi,\pm\pi)$ vector are decoupled, and thus are exclusively included in the Hamiltonian.

\begin{figure*}[t]
\hspace{2.cm}
\rotatebox[origin=c]{0}{\includegraphics[width=1.5\columnwidth]{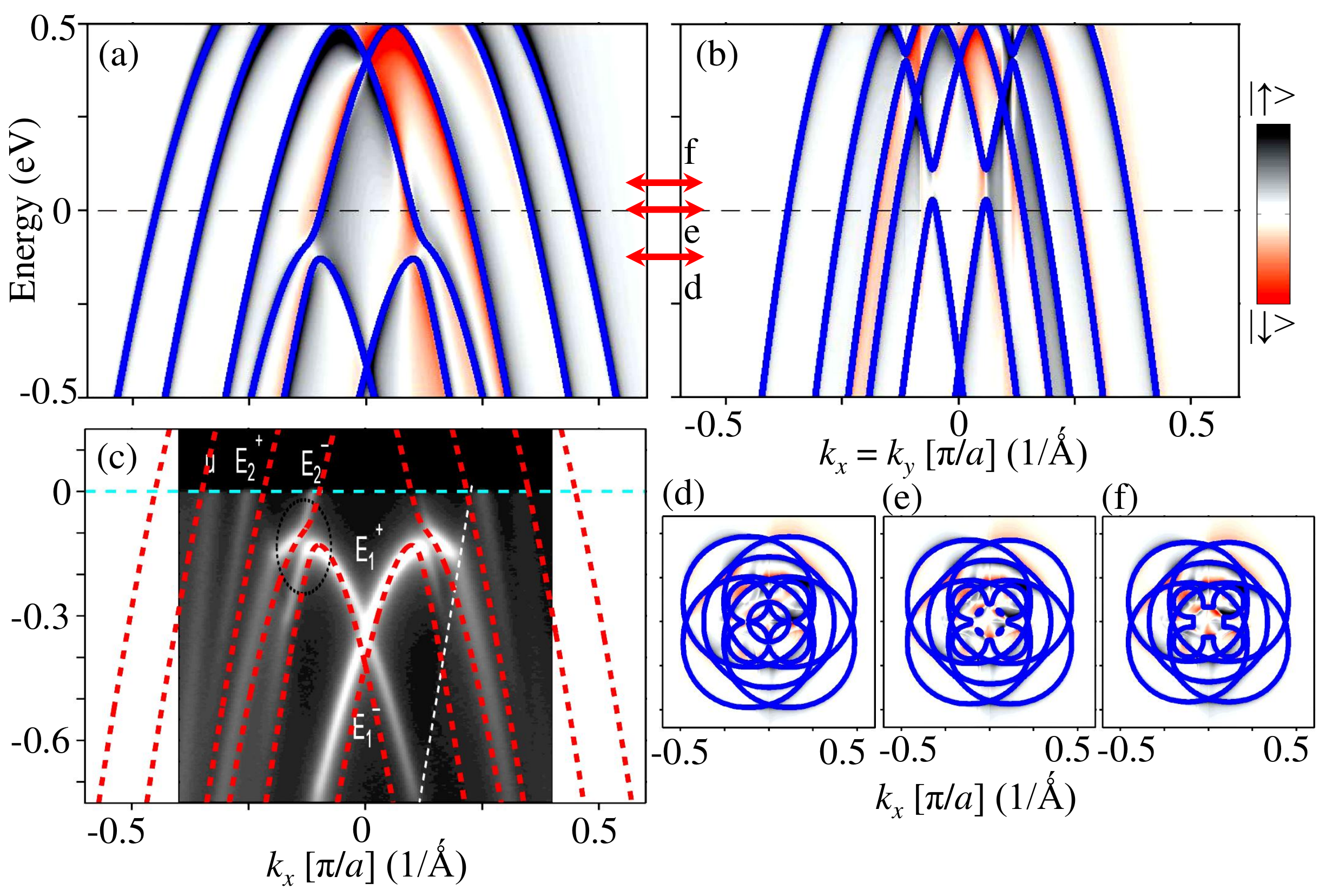}}
\caption{(Color online) Quasiparticle gapping due to SODW. ({a}) Electronic structure and associated spin-texture are plotted along (100)-direction. ({b}) Same as ({a}) but along (110) axis. The gap opening occurs below and above the Fermi level in both cases, while the spin-degeneracy at $\Gamma$-point is intact. Three horizontal arrows dictate three energy scales where the constant energy maps are presented in ({d}), ({e}), and ({f}). (c) The ARPES result of the gap opening in the BiAg$_2$ surface state (obtained via deposition of the sample on varying number of monolayers of Ag/Au heterostructure), taken from Ref.~\cite{BiAg2PRL}. The present theory, replotted from ({a}) on top of the experimental data with red dashed line, is in detailed agreement with the ARPES data. The number of sub-bands developed in the SODW is clearly evident in the experimental data as well, demonstrating further the interaction origin of the gap. Results are presented in the unfolded Brillouin zone to facilitate comparison with experimental data.}
\label{fig2}
\end{figure*}

{\it Spin-orbit density wave:-}Based on these results, we now desire to write down a two-body interacting Hamiltonian in the reduced Brillouin zone in which the Nambu-Gor'kov spinor becomes $\Psi_{\bm k}=[\psi_{{\bm k}\uparrow}, \psi_{{\bm k}\downarrow}, \psi_{{\bm k}+{\bm Q}\uparrow}, \psi_{{\bm k}+{\bm Q},\downarrow}]^T$. In this notation the interaction in the singlet-channel can be characterized by a contact interaction parameter $g$ and correspondingly, $H_I ({\bm k})=g\psi_{{\bm k},\uparrow}^{\dag}\psi_{{\bm k},\downarrow}\psi_{{\bm k}+{\bm Q},\uparrow}^{\dag} \psi_{{\bm k}+{\bm Q},\downarrow}$. In order to reduce the two-body problem into ordered Fermionic ground state, we decouple the interaction term $H_I$ by introducing a auxiliary spin-orbit field $\Delta ({\bm k})=g\psi_{{\bm k}+{\bm Q},\nu}^{\dag}[\sigma_z\otimes\sigma_x]_{\nu\nu^{\prime}}\psi_{{\bm k},\nu^{\prime}}$ [note that $\psi_{{\bm k},\nu}$ and $\psi_{{\bm k}+{\bm Q},\nu^{\prime}}$ belong to different helical states, having two Pauli matrices $\sigma_z$ and $\sigma_x$ for spin and momentum, and $\otimes$ represents a tensor product between them]. Employing mean-field approximation to the spin-orbit field $\Delta({\bm k}$), we obtain the total Hamiltonian as $H = \Psi_{\bm k}^{\dag}\left[H_0({\bm k}) + \Delta({\bm k})\sigma_z\otimes\sigma_x + h.c.\right]\Psi_{\bm k}$ which leads to the excitation spectrum as
\begin{eqnarray}
&E_{\bm k}^{\mu,\nu}=S_{+}^{\nu}+\mu [(S_{-}^{\nu})^2+\Delta^2]^{1/2},\nonumber\\
&{\rm with}\hspace{0.15cm}S^{\nu}_{\pm}({\bm k})=(E_{{\bm k},0}^{\nu}\pm E_{{\bm k},0}^{\bar{\nu}})/2,
\label{eq:eig}
\end{eqnarray}
where $\bar{\nu}=-\nu=\pm$ is the helical index due to SOC, and $\mu=\pm$ is the split band index due to translational symmetry breaking. We evaluate the order parameter $\Delta$ self-consistency as a function of temperature to a given contact potential. See supplementary material (SM)\cite{SM} for technical details. The obtained temperature dependence and critical temperature demonstrate the spontaneous behavior of mean-field gap opening.

{\it Quasiparticle gapping:-}Fig.~\ref{fig2} gives our main result in which the nature of the quasiparticle gap opening due to SODW is demonstrated and compared with angle-resolved photoemission spectroscopy (ARPES) data.\cite{BiAg2PRL} The blue lines are the coherent quasiparticle bands, plotted on top of the spin-resolved spectral weight [red (spin down) to black (spin up) colormap]. In the SODW state, two helical states split into several sub-bands, and a gap opens at the energy where two opposite helical states are connected by the ${\bm Q}$ vector. Along (100)-direction, the gap opening occurs in the filled state. The nature of gap opening and multiple number of shadow bands are in detailed agreement with recent ARPES data on the surface state of BiAg$_2$ alloy deposited on the monolayres (MLs) of Ag/Au(111) heterostructure.\cite{BiAg2PRB,BiAg2PRL} Similarly, along the diagonal direction, $\Delta$ appears slightly above the Fermi level, and tiny hole-like pockets develop, as visible on the FS map in Fig.~\ref{fig2}({e}). The constant energy surface map at an energy $E$=-110~meV illustrates how the main band and the folded band are nested along $(100)$-direction (in the particle-hole channel) where the quasiparticle gapping occurs, see Fig.~\ref{fig2}({d}). Similarly, at an energy $E$=62~meV above the Fermi level, the quasiparticle state is fully gapped along the diagonal direction; Fig.~\ref{fig2}({f}). 
%We emphasize that despite the reconstruction of the electronic states, the spin-degeneracy at $\Gamma$-point remains intact -- both in theory and experiment.

\begin{figure}[t]
%\hspace{2.cm}
\rotatebox[origin=c]{0}{\includegraphics[width=1.0\columnwidth]{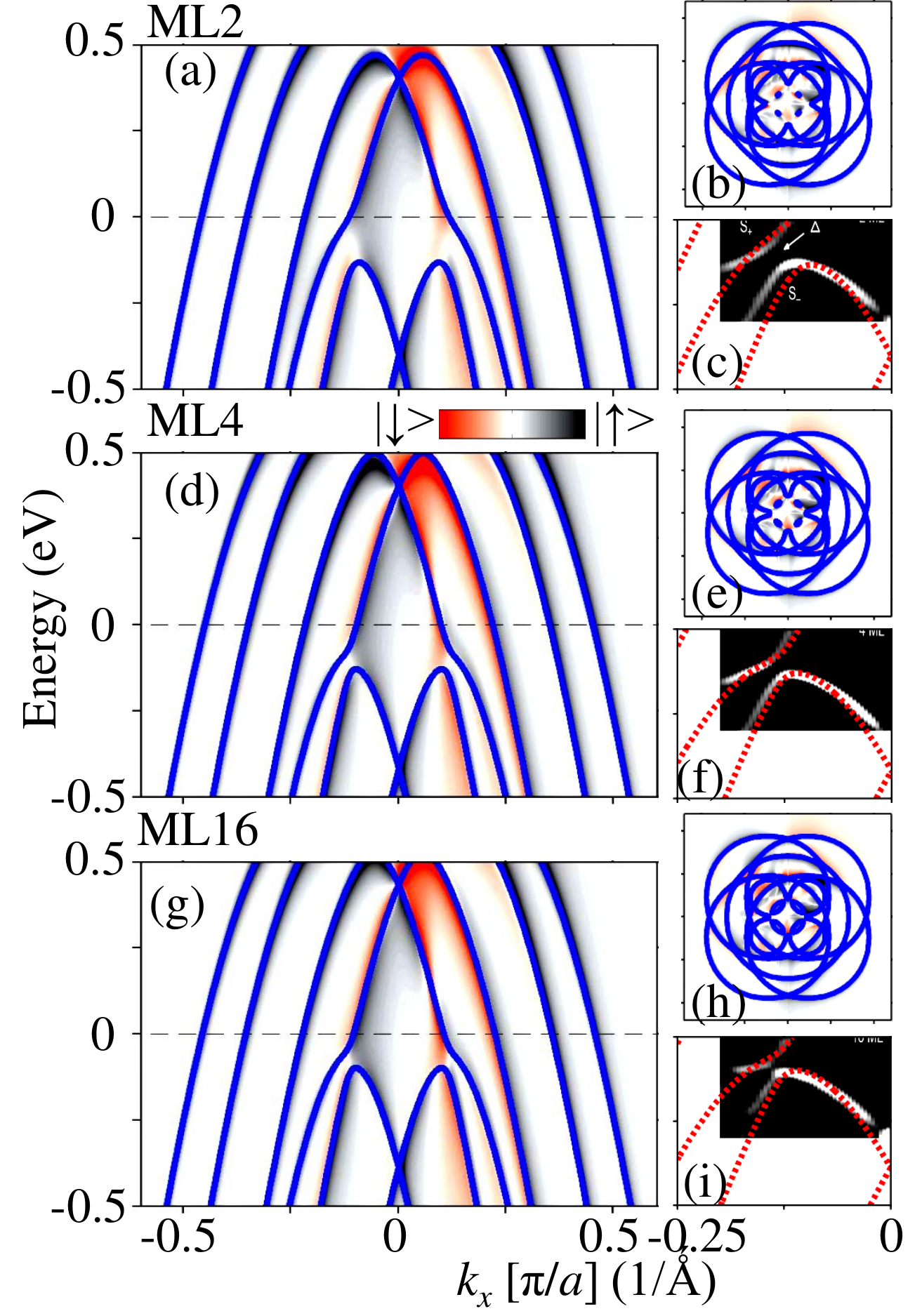}}
\caption{(Color online) Tunable spin-orbit order. ({a}) Quasiparticle dispersion along (100)-direction for self-consistently evaluated gap $\Delta$=120~meV. ({b}) Corresponding FS map. ({c}) Comparison with ARPES data for the case of ML2.\cite{BiAg2PRL} (d)-(f) Same as (a)-(c) but for a smaller gap of $\Delta=80$~meV and comparison with data for ML4 (reproduced from Fig.~\ref{fig2}). The result for even a smaller gap of $\Delta=60$~meV is compared with experimental data for ML16 in ({g})-({i}). As the gap gradually decreases with increasing number of monolayers of Ag/Au heterostructure, the location of the gap approaches the Fermi level, suggesting that the interaction induced spin-order effect is tunable, irrespective of a constant value of the non-interacting Rashba-coupling strength.}
\label{fig3}
\end{figure}

{\it Tunable spin-orbit order:-}
%For the experimental evidence of the spin-orbit order in 2D Fermi gas, we relate our theory to the quasiparticle gapping observed in a BiAg$_2$ surface alloy deposited on the monolayres (MLs) of Ag/Au(111) heterostructure.\cite{BiAg2PRB,BiAg2PRL} 
Interestingly, the ARPES data\cite{BiAg2PRL}  also reveals that the measured gap value $\Delta$ varies upon changing the thickness of the Ag film, with a clear evidence of Kramers degenerate point at $\Gamma$-momentum for all cases. These observations indicate that the systems undergoes quantum phase transition with a spontaneous symmetry breaking in the spin-orbit channel, whereas the time-reversal symmetry is not explicitly broken. It is interesting to note that a similar broken symmetry state other than time-reversal symmetry is obtained on the surface state of topological insulators,\cite{TISato,TIHasan} and also proposed to be responsible for the enigmatic `hidden-order' state in heavy fermion URu$_2$Si$_2$,\cite{DasHO} despite having different forms of spin-orbit coupling quantum state in these systems.

ARPES data plotted in Figs.~\ref{fig3}(c), \ref{fig3}(f), \ref{fig3}(i) show that the gap decreases from $\Delta=120$~meV for ML2 to $\Delta=80$~meV for ML4 to $\Delta=60$~meV for ML16. To theoretically explain this gap variation, we self-consistently tune the contact potential $g$, while keeping the Rashba term $\alpha_R$ fixed. We find that the gap closing is associated with chemical potential variation (beyond rigid band shift approximation), and the gapped region moves towards the Fermi level. We also find that the critical temperature below which the SODW sets in decreases with decreasing $g$; see SM\cite{SM}. Subsequently, the hole pockets along the diagonal directions grow in size. The large helical FSs with tunable area will be of considerable value for generating and detecting spin-resolved transports. In what follows, the new interaction induced spin-orbit effect is tunable and large, even when the intrinsic SOC is fixed and weak.

\begin{figure}[th]
%\hspace{2.cm}
\rotatebox[origin=c]{0}{\includegraphics[width=1.0\columnwidth]{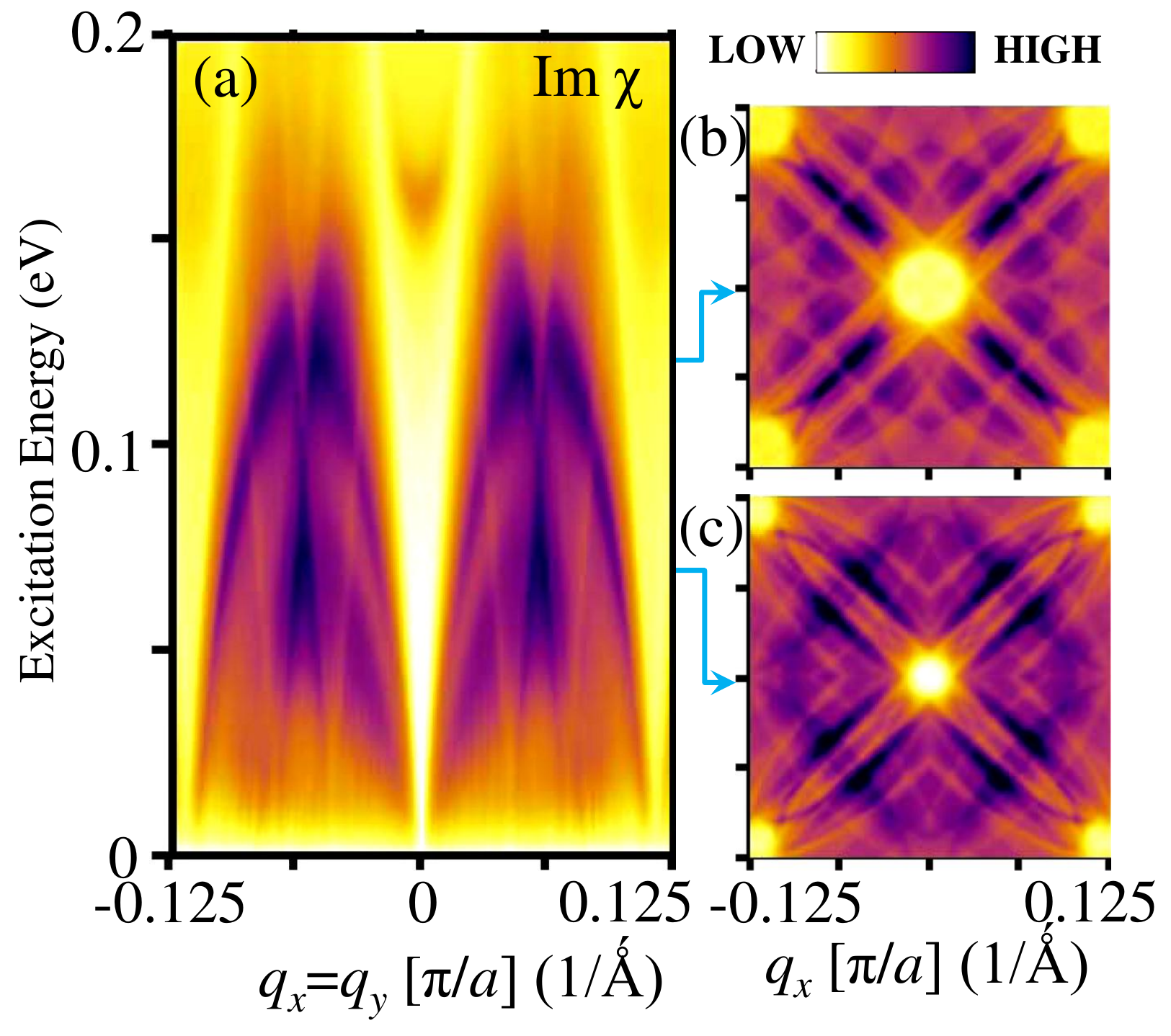}}
\caption{(Color online)  Collective spin-orbit mode. ({a}) Computed spectrum of imaginary part of RPA susceptibility in the spin-orbit channel in the SODW state plotted along (110) direction. The parameters for this calculation is same as deduced for ML2 configuration before. A sharp peak in intensity at ${\bm Q}\sim 0.115(\pi,\pi)$ is clearly visible around $\omega\sim\Delta\approx0.12$~eV. A weak downward dispersion branch away from this mode can be marked. The second mode around $\Delta/2$ is a multiband feature. ({b})-({c}) Constant energy profiles of the same susceptibility at first and second mode energies, respectively.}
\label{fig4}
\end{figure}

{\it Spin-orbit correlation function and emergent collective mode:-}Finally, to unravel the mechanism of SODW order, we deduce the emergence of associated collective excitations in the spin-orbit channel. The general form of the spin-orbit polarization vector is defined as $J({\bm q},\tau)=\sum_{{\bm k}}\psi^{\dag}_{{\bm k},\nu}(\tau)[\sigma_z\otimes\sigma_x]_{\nu\nu^{\prime}}\psi_{{\bm k}+{\bm q},\nu^{\prime}}(\tau)$, and correspondingly its bare correlator $\chi_0({\bm q},\tau)=\langle T_{\tau} J({\bm q},\tau)J(-{\bm q},0)\rangle$, where $\tau$ is the imaginary time and the operator $T_{\tau}$ denotes standard time-ordering between the fermionic fields. The result is presented as a function of excitation energy after performing Fourier transformation, and including many-body correction within the treatment of random-phase approximation (RPA): $\chi=\chi_0/(1-U\chi_0)$, where $U$ is the interaction matrix defined in the SM\cite{SM}.

The corresponding result of the imaginary part of RPA susceptibility $\chi$ is given in Fig.~4. Along the diagonal direction, a robust collective mode is visible at ${\bm Q}\sim0.115(\pi,\pi)$ with its energy proportional to the gap value. The result is presented for the case of ML2, however, the $\omega_{res}\sim\Delta$ relation is reproduced for other cases (in the bare susceptibility level, the peak appears at the gap energy, however, the many body RPA correction shifts the mode energy to a slightly lower value). A downward dispersion of the mode is also visible centering ${\bm Q}$, and extending to ${\bm q}\rightarrow 0$ and ${\bm q}\rightarrow 0.124(\pi,\pi)$ with vanishing intensity. A second collective mode appears at an energy which is about half of the first mode energy due to multi-band effect. The first mode is a robust feature tied to the emergent spontaneous symmetry breaking SODW, whereas the intensity and energy of the second mode varies and its appearance is subject to the strength of interaction. The constant energy profiles at the first and second mode energies are depicted in Figs.~4({b}),({c}). The bosonic spin-orbit excitation represents electron-electron interaction with simultaneous spin and orbital flips, or an entangled helical index flip.\cite{SM} A polarized inelastic neutron scattering measurement, which directly probes the imaginary part of the susceptibility, will be able to detect this mode. We note that no such collective mode develops in the charge susceptibility, and thus confirming that the present spin-orbit interaction does not activate any charge excitation.

{\it Outlook and discussion:-}The proposed spin-orbit order is a novel quantum phase of matter which can also arise in variety of other materials in which the non-interacting wavefunction is defined by exotic quantum number such as helical index, pseudospin, total angular momentum (typically in heavy-fermions and actinides) owing to SOC of various natures. Depending on the nature of the interaction and broken symmetry, the spin-orbit order can as well emerge as short range order in which other interacting phases such as quantum spin-Hall effect, spin-orbit nematic phase may exist. Here, we take a 2D electron gas as a representative example of how a SODW can arise due to the FS instability in a Rashba-type SOC background. In this context, it is interesting to point out the recent experimental findings of quasiparticle gapping in the surface state of topological insulator due to quantum phase transition. Such gap opening in the absence of time-reversal symmetry breaking and without the destruction of bulk topological properties,\cite{TISato,TIHasan} violates the conventional topological paradigm and criterion.\cite{TIKane} We envisage that it is suggestive of an emergent time-reversal invariant spin-orbit order, whose detail is required to be formulated in future study.

Some important advantages of the present spin-orbit order than the typical single-particle SOC can be noted. (1) In topological insulators or in Rashba systems, the functional use of spin-orbit effect is subject to counter propagating helical spin state which is topologically protected by Kramers spin-degeneracy with {\it zero} gap at the $\Gamma$-point.\cite{TIKane} Therefore the presence of any type of time-reversal breaking impurity or defects, with strength as small as infinitesimal value will destroy the protection, and thus barring the exciting usefulness of topologically protected transport properties. On the other hand, the present spin-orbit order is not only protected by symmetry, but most importantly it is thermodynamically shielded with a condensation energy equal to the tunable and large gap. Therefore, it would require a sufficiently large value of magnetic field such that its associated Zeeman energy $E_Z\sim g\mu_B B$ ($g$ is `g'-factor, $\mu_B$ is Bohr magneton)] can overcome the condensation gap energy. (2) Spin de-phasing time $\tau_s$ which is an important ingredient for the spintronics and quantum computing applications is considerably larger here (determined by $E_Z$ energy), and is tunable. (3) Another crucial benefit of the interaction induced spin-orbit effect is that, unlike in typical electric field or magnetic field induced SOC, it does not necessarily activate a charge flow, and thus will be highly valuable for solely generating spin-transport.

Another experimental verification of the broken symmetry spin-orbit order is detection of spin Nernst effect. Since Nernst effect is sensitive to reconstructed FS, a manifestation of broken symmetry phase,\cite{Nernst,DasNernst} spin-orbit order will generate a spin-resolved thermal current which are detectable in recent days laboratory facilities.\cite{spinNernst}

\begin{acknowledgments}
The author is grateful to M. J. Graf, A. V. Balatsky, J.-X. Zhu, S. Raghu, P. W\"olfle for many fruitful discussions. The work is supported by the U.S. DOE through the Office of Science (BES) and the LDRD Program and faciliated by NERSC computing allocation.
\end{acknowledgments}

%\input{BdG_Pu.bbl}
%\end{document}

\section{Supplementary Material}

In the supplementary mayerial we provide the detailed derivation of the spin-orbit order parameter. Due to the definite chirality of the FS, different orientations of ${\bm Q}_i=0.115(\pm\pi,\pm\pi)$ are decoupled and are exclusively included in the Hamiltonian. In this spirit we define the Nambu-Gor'kov spinor as $\Psi_{\bm k}=[\psi_{{\bm k},\uparrow}, \psi_{{\bm k},\downarrow}, \psi_{{\bm k}+{\bm Q}_i,\uparrow}, \psi_{{\bm k}+{\bm Q}_i,\downarrow}, ...]^T$, where $i=1-4$. Since macroscopically the gap value is same for all four ${\bm Q}_i$ vectors, the interaction potential is constant which gives the two-body spin-orbit term as $H_I ({\bm k})=g\sum_i\psi_{{\bm k},\uparrow}^{\dag}\psi_{{\bm k},\downarrow}\psi_{{\bm k}+{\bm Q}_i,\uparrow}^{\dag} \psi_{{\bm k}+{\bm Q}_i,\downarrow}$. In the functional path integral formalism, the partition function of the Fermi gas is $\mathcal{Z}=\int {\mathcal D}{\Psi^{\dag}}{\mathcal D}{\Psi}~\exp{(-S[\Psi^{\dag},\Psi])}$ ($\hbar=k_B=1$ throughout), where the action is $S[\Psi^{\dag},\Psi]=\int_0^{\beta}d\tau\sum^{\prime}_{\bm k} [\Psi^{\dag}\partial_\tau \Psi + H_0 + H_I]$, with $\beta=1/T$ and the prime over the summation represents that the summation is performed in the reduced Brillouin zone.

In order to reduce the above-derived action into an ordered Fermionic problem, we decouple the four-field interaction term by introducing the auxiliary spin-orbit coupling field $\Delta ({\bm k})=g\psi_{{\bm k}+{\bm Q},\nu}^{\dag}[\sigma_z\otimes\sigma_x]_{\nu\nu^{\prime}}\psi_{{\bm k},\nu^{\prime}}$. Employing the Hubbard-Stratonovich transformation\cite{HubbardStrat} and integrating out the fermionic fields, we obtain $\mathcal{Z}=\int \mathcal{D}\Delta^{\dag}\mathcal{D}\Delta\exp{(-S_{eff}[\Delta^{\dag},\Delta])}$, where the effective spin-orbit order reads $S_{eff}[\Delta^{\dag},\Delta]=-\frac{1}{2}{\rm Tr}\ln{[G^{-1}({\bm k},\tau)]}-\int_{0}^{\beta}\sum_{\bm k}d\tau |\Delta({\bm k})|^2/g$. The inverse single-particle Green's function in the fermionic Matsubara frequency is obtained as
\begin{eqnarray}
G^{-1}=\left(
\begin{array}{ccc}
% line 1
i\omega_n{\bf 1}-H_0({\bm k}) & {\bm \sigma}_x\Delta&\ldots\\
% line 2
{\bm \sigma}_x\Delta^{\dag} & i\omega_n{\bf 1}-H_0({\bm k}+{\bm Q}_1)&\ldots\\
\vdots&\vdots&\ddots
\end{array}
\right).
%%%%%%%%%%%%
\label{greenfn}
\end{eqnarray}
The other terms belong to the three values of ${\bm Q}_i$. At the mean-field level, $\Delta({\bm k})$ represents the gap parameter which is same for all values of ${\bm Q}_i$. In this case, the effective action further simplifies to $S_{eff}[\Delta^{\dag},\Delta]=-\frac{1}{2}\ln{[{\rm det}G^{-1}({\bm k},i\omega_n)]}-N\beta |\Delta({\bm k})|^2/g$, where $N$ is the size of the system. From $ {\rm det}[G^{-1}({\bm k},E)]=0$, we derive the excitation spectrum as given in Eq.~1 in the main text. Finally, by using the relation for the thermodynamic potential $\Omega=-1/\beta \ln{\mathcal{Z}}$, we obtain the self-consistent condition for the number operator and the gap function as:
\begin{eqnarray}
n_{\uparrow/\downarrow}&=&\sum_{{\bm k},\nu}^{\prime}\int_{-\infty}^{\infty}\frac{d\omega}{2\pi} {\rm Im} G_{11/22}({\bm k},\nu,\omega+i\delta)f(\omega),
\label{Eq:n}\\
 \Delta&=&g\sum_{{\bm k},\nu}^{\prime}\int_{-\infty}^{\infty}\frac{d\omega}{2\pi} \nu {\rm Im} G_{12}({\bm k},\nu,\omega+i\delta)f(\omega),
\label{Eq:gap}
\end{eqnarray}
 where $\delta$ is infinitesimally small broadening parameter as $f(\omega)$ is the Fermi-Dirac distribution function for the excitation energy of $\omega$. We solve Eqs.~\ref{Eq:n} and \ref{Eq:gap}, self-consistently to obtain the value of quasiparticle gap to match the experiment.

\begin{figure}[t]
%\hspace{0.cm}
\rotatebox[origin=c]{0}{\includegraphics[width=0.85\columnwidth]{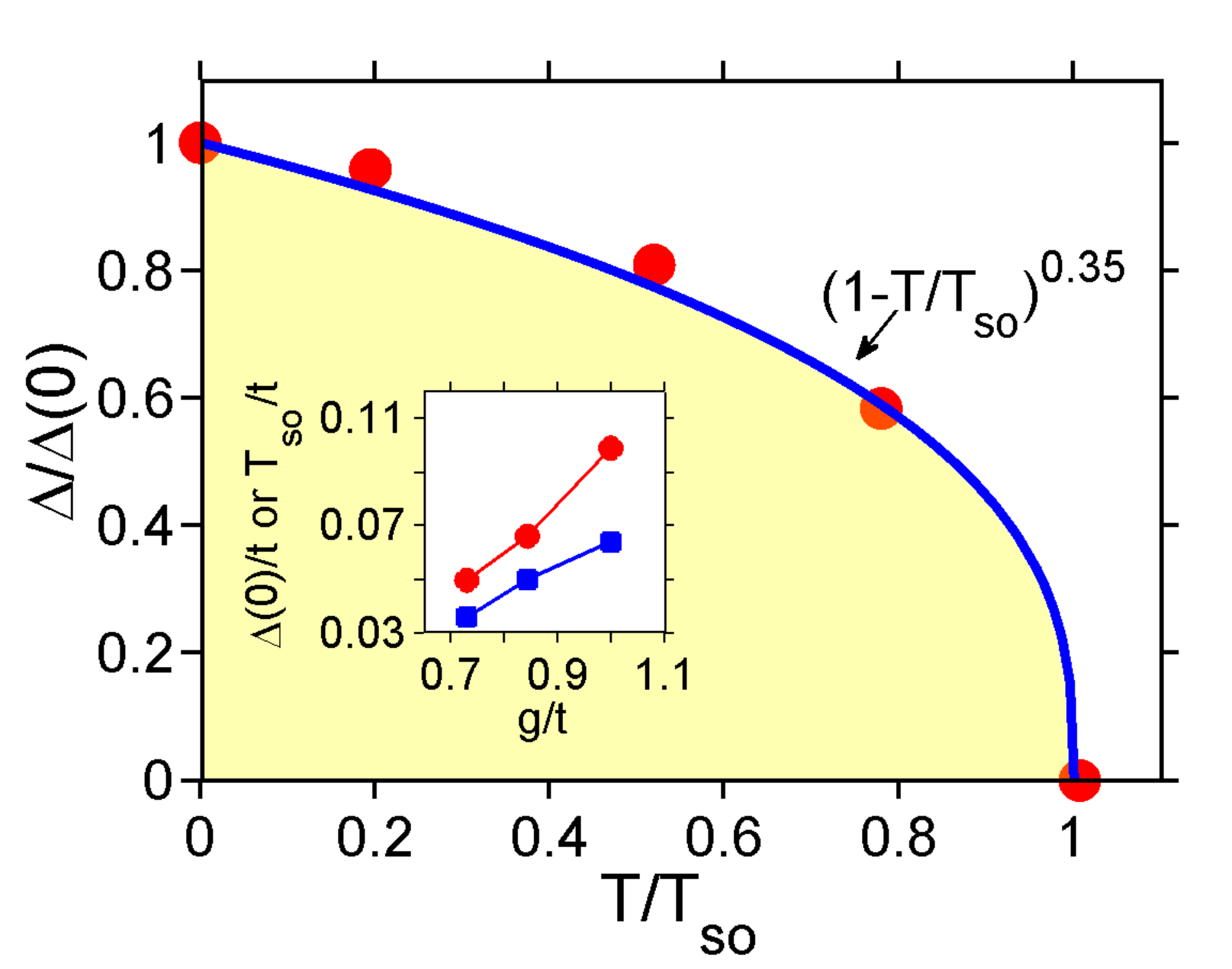}}
\caption{Self-consistent gap values as a function of temperature for $\Delta(0)/|t|\approx$0.1 with $g/|t|$=1. The critical temperature for the spin-orbit density wave, $T_{so}$, is estimated for this case to be around 800~K. The line depicts a fit of the function $\Delta(T)/\Delta(0)=(1-T/T_{so})^{0.35}$, which is close to the BCS critical exponent of 0.5. {\it Inset:} The critical temperature $T_{so}$ (blue squares) and gap amplitude $\Delta(0)$ (red circles) varies linearly with the interaction potential strength $g$. The data are presented with respect to $|t|$.}
\label{sfig1}
\end{figure}

The values of all parameters are obtained by fitting the quasiparticle-spectrum and gap with the experimental value from Ref.~\cite{BiAg2PRL}. We find $t=-1.216$~eV, and $\alpha_R=1.25/|t|$. The interaction potential is self-consistently evaluated to be $g/|t|$=1, 0.84, and 0.73 which gives the experimental gap values of $\Delta=$0.12, 0.08 and 0.06~eV for ML2, ML4 and ML16, respectively. We also deduce the temperature dependence of the gap parameter as given in the supplementary Fig.~\ref{sfig1}. The result reproduces a mean-field like behavior with the critical exponent $0.35$, which can be compared with the typical BCS value of $0.5$. The critical temperature below which a spin-orbit order sets in relies on the contact potential (see inset to Fig.~\ref{sfig1}).

\section{Spin-orbit excitation}

Finally, we expand on the deduction of the spin-orbit susceptibility. The spin-orbit operator is defined in the main text as
\begin{eqnarray}
J({\bm q},\tau)=\sum_{{\bm k}}\psi^{\dag}_{{\bm k},\nu}(\tau)[\sigma_z\otimes\sigma_x]_{\nu\nu^{\prime}}\psi_{{\bm k}+{\bm q},\nu^{\prime}}(\tau).
\end{eqnarray}

Substituting the expression for spin-orbit polarization in the spin-orbit susceptibility formula, we obtain
\begin{widetext}
\begin{eqnarray}
\chi_0({\bm q},\tau)&=&-\frac{1}{N}\left\langle T_{\tau}J({\bm q},\tau)J(-{\bm q},0)\right\rangle
=-\frac{1}{N}\sum_{{\bm k},{\bm k}^{\prime}}\langle T_{\tau} \psi^{\dag}_{{\bm k},\nu}(\tau)\psi_{{\bm k}+{\bm q},\nu^{\prime}}(\tau)\psi^{\dag}_{{\bm k}^{\prime},\nu}(0)\psi_{{\bm k}^{\prime}-{\bm q},\nu^{\prime}}(0)\rangle\nonumber\\
\end{eqnarray}
\end{widetext}
Rewriting the above susceptibility in the reduced Brillouin zone (denoted by prime over the summation), we get
\begin{widetext}
\begin{eqnarray}
\chi_0({\bm q},\tau)&=&-\frac{1}{2N}\sum^{\prime}_{{\bm k},{\bm k}^{\prime}}\left[\langle T_{\tau} \psi^{\dag}_{{\bm k},\nu}(\tau)\psi_{{\bm k}+{\bm q},\nu^{\prime}}(\tau)\psi^{\dag}_{{\bm k}^{\prime},\nu}(0)\psi_{{\bm k}^{\prime}-{\bm q},\nu^{\prime}}(0)\rangle 
+\sum_i\left(\langle T_{\tau} \psi^{\dag}_{{\bm k},\nu}(\tau)\psi_{{\bm k}+{\bm q},\nu^{\prime}}(\tau)\psi^{\dag}_{{\bm k}^{\prime}+{\bm Q}_i,\nu}(0)\psi_{{\bm k}^{\prime}-{\bm q}+{\bm Q}_i,\nu^{\prime}}(0)\rangle\right.\right.\nonumber\\
&&~~~\left.\left.+\langle T_{\tau} \psi^{\dag}_{{\bm k}+{\bm Q}_i,\nu}(\tau)\psi_{{\bm k}+{\bm q}+{\bm Q}_i,\nu^{\prime}}(\tau)\psi^{\dag}_{{\bm k}^{\prime},\nu}(0)\psi_{{\bm k}^{\prime}-{\bm q},\nu^{\prime}}(0)\rangle
+\langle T_{\tau} \psi^{\dag}_{{\bm k}+{\bm Q}_i,\nu}(\tau)\psi_{{\bm k}+{\bm q}+{\bm Q}_i,\nu^{\prime}}(\tau)\psi^{\dag}_{{\bm k}^{\prime}+{\bm Q}_i,\nu}(0)\psi_{{\bm k}^{\prime}-{\bm q}+{\bm Q}_i,\nu^{\prime}}(0)\rangle\right)\right]\nonumber\\
\end{eqnarray}
\end{widetext}

\begin{figure}[h]
%\hspace{0.cm}
\rotatebox[origin=c]{0}{\includegraphics[width=0.85\columnwidth]{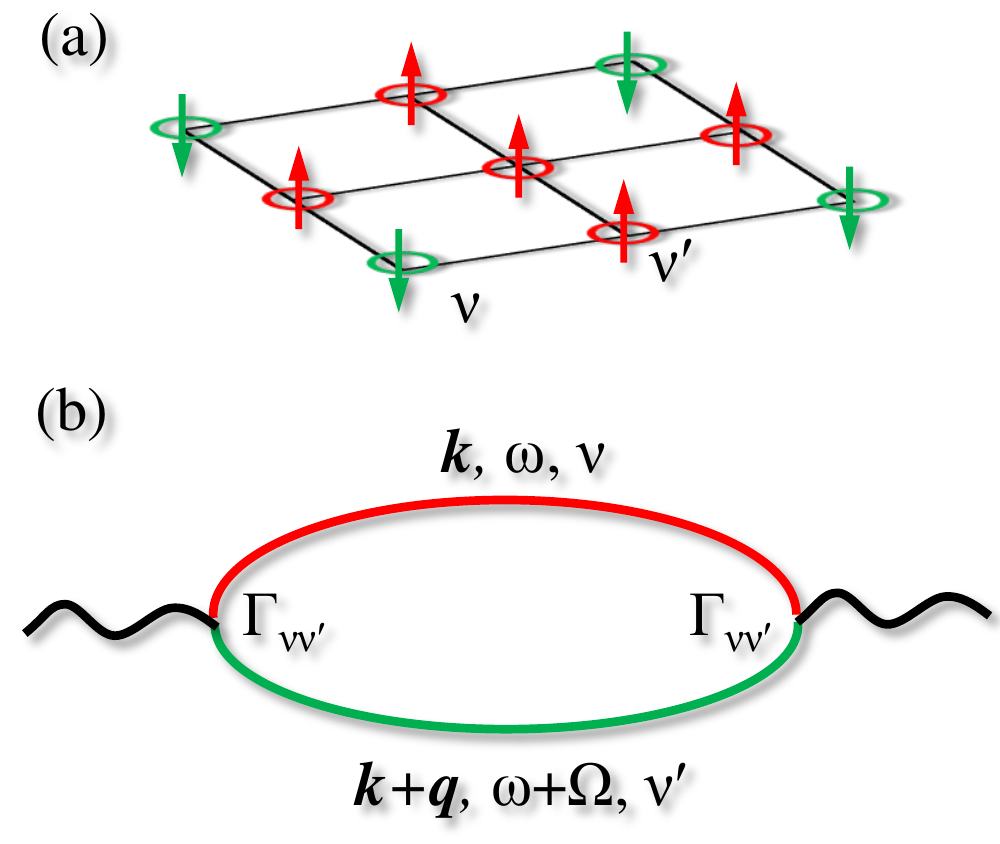}}
\caption{(a) Real-space view of spin-orbit density wave having two opposite spin-orbit states sitting in different sublattices. Arrows stand for spin while circles with different colors give different orbitals. We chose a commensurate modulation here along the diagonal for illustration convenience. (b) Spin-orbit bubble which is responsible for electron-electron interactions having opposite spin and orbital quantum numbers. The interaction vertex $\Gamma_{\nu\nu^{\prime}}=[\sigma_z\otimes\sigma_x]_{\nu\nu^{\prime}}$.}
\label{sfig2}
\end{figure}

Expanding the above four-fermionic fields into two ordered fields in the spin-orbit channel and employing the Wick's theorem,\cite{wickstheory} we can write down the above equation in terms of Green's functions as
\begin{eqnarray}
\chi_0({\bm q},\tau)&=&-\frac{1}{2N}\sum_{\bm k}^{\prime}\sum_{ij}G_{ij}({\bm k},\nu,\tau)G_{ij}({\bm k}+{\bm q},\nu^{\prime},-\tau), \nonumber\\
\end{eqnarray}
where $G_{ij}$ are the components of the Green's function given in Eq.~1 above. Here we employ the momentum conservation relation ${\bm k}^{\prime}={\bm k}+{\bm q}$ or ${\bm k}^{\prime}={\bm k}+{\bm q}+{\bm Q}_i$. Each term in $\chi_0$ represent a polarization bubble as represented in supplementary Fig.~\ref{sfig2}(b) which is responsible for interaction between two electrons with opposite orbitals and spins, or with opposite helical index $\nu$. The explicit form of the susceptibility tensor is $\chi_0^{ijkl}({\bm q},ip_m)=-1/N\sum_{{\bm k},n,\mu,\mu^{\prime}}\phi_{\mu}^{i\dag}({\bm k})\phi_{\nu}^j({\bm k})\phi_{\mu^{\prime}}^{k\dag}({\bm k}+{\bm q})\phi_{\mu^{\prime}}^l({\bm k}+{\bm q})(f(E_{\bm k}^{\mu})-f(E_{{\bm k}+{\bm q}}^{\mu^{\prime}})/(i\omega_n-E_{\bm k}^{\mu}-E_{{\bm k}+{\bm q}}^{\mu^{\prime}}).$ Here $\phi_{\mu}^{i}$ and $E^{\mu}$ are the eigenvector and eigenvalue of the Hamiltonian deduced in Eq.~\ref{greenfn}, with $\mu,\mu^{\prime}$ being the corresponding band indices, and $i,j,k,l$ their orbital counterparts. Within RPA framework, the perturbation interaction term for the susceptibility is $H_{RPA}=U\sum_{\nu}n_{\nu}n_{\nu}+U^{\prime}\sum_{\nu}n_{\nu}n_{\bar{\nu}}$, where $\nu=-\bar{\nu}=\pm$ is the helical index and $U$, and $U^{\prime}$ are the interaction strength between same and opposite helical states. Since the interaction here is Coulomb type which acts on charge particles, therefore, without any loss of generality, we take $U=U^{\prime}\sim g=0.35$~eV for ML2 configuration. It is interesting to note here that despite the same value of interaction between both helical states, the long-range order is prohibited along the spin-channel due to strong spin-orbit coupling ($U<<\alpha_R$), but from symmetry argument, an order parameter in the spin-orbit channel ($U^{\prime}$-term) is allowed.


\begin{thebibliography}{10}
\bibitem{parameters} The values of all parameters are obtained by fitting the quasiparticle-spectrum and gap with the experimental value from Ref.~\onlinecite{BiAg2PRL}. We find $t=-1.216$~eV, and $\alpha_R=1.25/|t|$. The interaction potential is self-consistently evaluated to be $g/|t|$=1, 0.84, and 0.73 which gives the experimental gap values of $\Delta=$0.12, 0.08 and 0.06 in eV unit for ML2, ML4 and ML16, respectively.
\end{thebibliography}
\end{document}